\documentclass[sigconf]{acmart} 
\AtBeginDocument{%
  \providecommand\BibTeX{{%
    \normalfont B\kern-0.5em{\scshape i\kern-0.25em b}\kern-0.8em\TeX}}}

\copyrightyear{2023} 
\acmYear{2023} 
\setcopyright{rightsretained} 
\acmConference[IDC '23]{Interaction Design and Children}{June 19--23, 2023}{Chicago, IL, USA}
\acmBooktitle{Interaction Design and Children (IDC '23), June 19--23, 2023, Chicago, IL, USA}
\acmDOI{10.1145/3585088.3589356}
\acmISBN{979-8-4007-0131-3/23/06}

\acmConference[IDC '23]{Collaborative Machine Learning Model Building with Families Using Co-ML}{June 19--22,
  2023}{}

\usepackage{array}

\newcommand{\todo}[1]{{\textcolor{red}{[#1]}\normalfont}}

\usepackage{xspace,xpunctuate}
\usepackage{enumitem}
\usepackage{subfig}

\newcommand{\system}{Co-ML\xspace}

\begin{document}

\title{Collaborative Machine Learning Model Building with Families Using Co-ML}


\author{Tiffany Tseng, Jennifer King Chen, Mona Abdelrahman, Mary Beth Kery, Fred Hohman, Adriana Hilliard, R. Benjamin Shapiro}
\email{{tiffanytseng, jennkingchen, mona_abdelrahman, mkery, fredhohman, adri, nerd}@apple.com}
\affiliation{
  \institution{Apple}
  \city{Seattle}
  \state{WA}
  \country{USA}
  }

% \renewcommand{\shortauthors}{Tseng et al.}

\begin{abstract}
Existing novice-friendly machine learning (ML) modeling tools center around a solo user experience, where a single user collects only their own data to build a model. However, solo modeling experiences limit valuable opportunities for encountering alternative ideas and approaches that can arise when learners work together; consequently, it often precludes encountering critical issues in ML around data representation and diversity that can surface when different perspectives are manifested in a group-constructed data set. To address this issue, we created Co-ML – a tablet-based app for learners to collaboratively build ML image classifiers through an end-to-end, iterative model-building process. In this paper, we illustrate the feasibility and potential richness of collaborative modeling by presenting an in-depth case study of a family (two children 11 and 14-years-old working with their parents) using Co-ML in a facilitated introductory ML activity at home. We share the Co-ML system design and contribute a discussion of how using Co-ML in a collaborative activity enabled beginners to collectively engage with dataset design considerations underrepresented in prior work such as data diversity, class imbalance, and data quality. We discuss how a distributed collaborative process, in which individuals can take on different model-building responsibilities, provides a rich context for children and adults to learn ML dataset design.
\end{abstract}




\begin{teaserfigure}
\centering
  \includegraphics[width=0.95\linewidth]{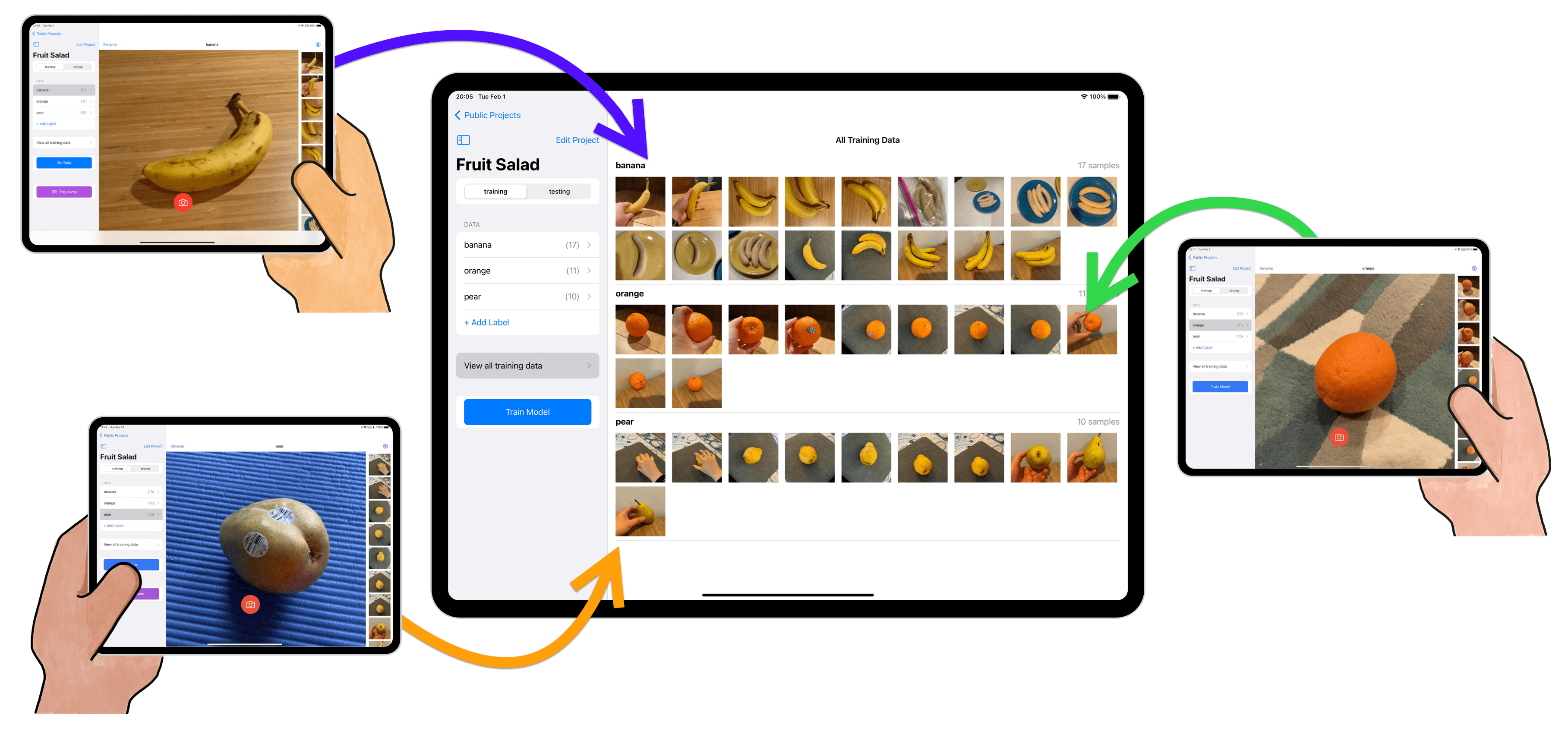}
  \caption{\system tablet app for collaborative ML model building. Using Co-ML, multiple users contribute data to a shared dataset and work together to train and test their own image classifiers.}
  \Description{An iPad showing a shared training data set surrounded by individual ipads adding photographs of training data.}
  \label{fig:teaser}
\end{teaserfigure}

\maketitle

\section{Introduction}
\label{sec:intro}
A pressing issue in ML today is the design of responsible AI systems that minimizes bias and works well in diverse contexts. Responsible ML systems require rich, varied data representative of the range of scenarios in which a model may be applied. A lack of balanced datasets has enormous public consequences, such as systems that do not account for differences in skintone and gender \cite{buolamwini2018gender} or are grounded in historical data with embedded inequalities, and in turn reproduce those inequalities \cite{eubanks2018automating, noble2018algorithms}. Teaching best practices for creating balanced datasets is an essential first step towards promoting critical conversations and education about ML systems. Yet with K-12 ML education being relatively nascent \cite{ai4k12}, best practices for teaching dataset design are still unestablished, and research is needed on how these ideas can be embedded into ML learning experiences and tools for beginners.

Existing ML modeling tools for novices like Teachable Machine \cite{carney2020teachable} limit opportunities for encountering dataset diversity issues because they center around a solo modeling experience, where a single user typically collects data and tests models using only their own data. Solo experiences can result in imbalanced datasets or otherwise reflect an individual's own perspective or unconscious biases. In contrast, we posit that a collaborative modeling experience, in which multiple people contribute to data collection and model testing, can facilitate the creation of balanced datasets representing multiple points of view.

To understand the role collaboration can play in learning balanced dataset design, we developed Co-ML, a tablet-based app for groups to build image classifiers. With Co-ML, groups decide on items for their classifier to recognize, and individuals collectively add to a shared dataset by adding images from their own devices. They then work together to train an ML model, test its performance, and iterate on their model by refining the underlying data.

In this paper, we describe how we designed Co-ML to support collaborative model building and promote good dataset design practices, via features like a shared dashboard for reviewing data, an interface for adding and evaluating test data, as well as a game that engages users in playfully and systematically testing model performance. We designed a 2-hour facilitated activity for families to build ML models with Co-ML using everyday items at home. To illustrate the affordances of the collaborative modeling experience and the ML dataset design ideas users encountered, we share an in-depth case study of a single family (mother and father working alongside their two sons, ages 11 and 14-years old) using Co-ML. Through this case study, we report on how the collaborative modeling experience supported by Co-ML provided a rich context to discuss, debate, and test theories about factors in dataset design critical to model performance.

\section{Related Work}
\label{sec:related-work}
We begin with a background on ML concepts and practices that underlie model building and are central to how models are created with Co-ML. Next, we summarize related work on AI literacy for children and collaborative social learning, highlighting opportunities for collaborative model building to support education on dataset design best practices.

\subsection{ML Model Building}
Machine Learning is a type of AI in which computers detect patterns in data to make useful predictions. \textit{Image classifiers} are a type of ML model that classifies unseen data such as images into predetermined classes or labels, based upon those learned patterns. In \textit{supervised} ML, classifiers are first trained on human-labeled data (\textit{training data}) using one of many possible underlying algorithms. The result of training is a \textit{model} that takes input (new data) and returns an output (a classification of what the new data is). When training a model, an algorithm detects a set of \textit{features}, or characteristics of the data, to distinguish between labels; then when testing a classifier on new data, the model returns a probability distribution for its \textit{confidence} in associating a given label with each element of the new data. For example, to build an ML classifier to identify fruits, a model would consist of labels for each type of fruit (e.g., banana, orange, and pear). Training data would consist of \textit{samples}, or images of each fruit labelled with the type of fruit pictured. Collecting these images is a process called \textit{data collection}. When testing the model (\textit{model testing}), human modeler(s) assess how accurately the model classifies new data. 

A substantial part of the ML model-building pipeline focuses on data, including its collection, cleaning, and analysis. It is often easier to improve model performance by iterating on data, such as adding more samples, rather than tweaking model architecture or algorithms \cite{hohman2020understanding}. When constructing quality datasets, ML practitioners need to consider how \textit{representative} the data is (that it accurately reflects characteristics of the labels), how \textit{diverse} it is (that it presents a variety of use cases and contexts), and how \textit{balanced} it is (ideally, the distribution of samples across labels is roughly equal). Model failures often result from data challenges. For example, commercial facial recognition algorithms that worked poorly for people of color \cite{buolamwini2018gender}, an issue known as \textit{class imbalance}. Having many users collecting data often contributes to more diverse datasets \cite{hong2020crowdsourcing} that can match a range of real-world use cases \cite{barbu2019objectnet}. Taking the same fruit classifier example, model designers might consider distinct features that distinguish labels, including representations such as color and shape. Data diversity might entail capturing fruit in different lighting conditions (indoor or outdoor), states (whole or cut), and camera angles (such as top-down or side views). If any images are found to be unrepresentative (such as being blurry or otherwise low quality, or being mislabeled), they can be removed or edited, a process referred to as \textit{quality control}. The dataset should have relatively equal number of samples of banana, oranges, and pears to avoid issues caused by class imbalance. Integrating dataset design practices into introductory ML experiences has the potential to prepare learners to understand pitfalls of biased models and ultimately inform their understanding of responsible ML design.

\subsection{AI Literacy for Youth}
\label{sec:related-work-ai-literacy}
AI4K12 \cite{ai4k12} is one of several efforts \cite{lee2021developing, long2020ai, zhou2020designing} to identify K-12 AI learning goals. AI4K12 has outlined a set of big ideas around ML, ranging from how computers perceive the world, how they represent data, and how they learn from data. While dataset design is one component of their framework (including helping learners examine features of training data, identify potential sources of bias, and understand how to properly balance datasets), more research is needed about how best to support youth learning these practices.

A growing set of tools enable youth to play with off-the-shelf AI services \cite{druga2018growing, kahn2017child} or even build models from scratch by collecting their own data and training a model (typically using visual programming or no-code options \cite{von2021visual}). Existing model-building tools for youth enable capturing physical activity \cite{agassi2019scratch, zimmermann2019youth, zimmermann2020youth} and building interactive toys that respond to custom gestures \cite{tseng2021plushpal}. Teachable Machine\cite{carney2020teachable} (inspired by Wekinator \cite{fiebrink2010wekinator}) supports beginners more broadly, with building ML models on the web and testing them in real time via a live classification interface.

However, a limitation of existing tools is that they center around solo model building, with limited opportunity to encounter dataset design issues such as a lack of diversity. This is because with existing tools, beginners largely build models that only work for themselves, using only their own data; this may lead to biased data and inaccurate model performance metrics since the ways a model is tested represent a single point of view. The value of different perspectives in model building has been demonstrated in an evaluation of Teachable Machine in which youth exchanged models with their peers, helping them identify limitations in how their model can work for others \cite{dwivedi2021exploring}. With growing public discourse around bias in AI systems, curricula have been developed to expose ethical concerns to young people \cite{payne2019ethics, williams2020train}. Creating tools that aid beginners in building balanced, diverse datasets is critical to support young people interpreting limitations of existing ML systems and empowering them to build ethical ML systems in the future; we believe that a collaborative tool in which multiple viewpoints are represented in the data can be especially fruitful towards this end.

\subsection{Social \& Family Learning}
The value of social learning, where individuals have the opportunity to learn with and from peers, is substantiated by decades of work in social psychology and child development. Our work is inspired by the foundational socio-constructivst idea that knowledge stems from socially mediated interactions \cite{vygotsky1978mind} in which collaboration supports the construction of knowledge through conversation and shared action, with learners interpreting information and using evidence to draw new conclusions collectively \cite{blumenfeld1996learning} and make arguments and explanations to one another \cite{berland2009making, roschelle1992learning}.  

Parents can play critical roles in fostering their children's interests and skill development in technology. Previous work has identified a range of roles parents take on, including project collaborator (working directly with their children) and learning broker (identifying resources for children to use) \cite{barron2009parents, roque2016m}. Activities that support fluent participation and flexible roles adaptive to a family's personal styles and learning agendas are valuable in a variety of contexts, including at-home enrichment \cite{barron2009parents}, workshops and camps \cite{lin2012investigation, roque2016m}, and museums \cite{zimmerman2010parents}. Regardless of their technical fluency, parents can support their children's learning through mediation strategies like posing questions and drawing attention \cite{zimmerman2010parents}. In programming activities, parent-child pairs have been found to produce more ``compact and well-structured'' code with fewer errors compared to children working alone; additionally, children reflected on their solutions more when working alongside a parent \cite{lin2012investigation}.  

Early work on families learning about AI has examined how families engage with concepts like semantic networks using paper-based games \cite{long2021role, long2022family}, studied how families interpret AI technologies like voice assistants \cite{beneteau2019communication}, and looked at the roles parents can play in facilitating children learning about AI \cite{druga2022family}. A distinction between the contributions of our research and prior work is that we center families learning AI through an experience where novices build functioning ML models together by iterating on the underlying dataset, empowering beginners to wrestle with ML dataset design issues firsthand. Thus, our focus is not on how families interact with existing production ML systems and how they believe they work, but rather how they go about building ML systems themselves and develop understandings of the role dataset diversity, features, and class balance work in how ML systems make decisions.
\section{Co-ML System}
\label{sec:system}

The Co-ML iPad app supports collaborative ML model-building through an end-to-end iterative workflow consisting of \textbf{defining an ontology of labels} for their classifier to identify, \textbf{collecting and reviewing training data} by taking photos on the tablet, \textbf{training a model} on device, \textbf{testing the model} on new data, \textbf{evaluating the model} by playing an in-app game, and \textbf{iterating on the model} by revising the underlying dataset.

We begin our design of Co-ML with image classification because of the interpretability and support for visual inspection compared to other modalities \cite{blackwell2001cognitive}. Modalities like sound may also be difficult to support in a co-located collaborative experience because of interference when multiple people speak simultaneously. Co-ML is optimized for tablets because the mobility of tablets reduces friction for collecting data in the wild (compared to laptops) while their larger screen size (compared to phones) can better support review and inspection of data. Computationally, tablets are powerful enough to generate and test ML models using small datasets (hundreds to thousands of images).

\subsection{Defining an Ontology of Labels}
First, users define a set of labels, or classes of objects a user wants the classifier to tell apart. Labels can be modified over time. In our example project used throughout this section, the labels are Banana, Orange, and Pear. 

\begin{figure*}
  \centering
  \includegraphics[width=\linewidth]{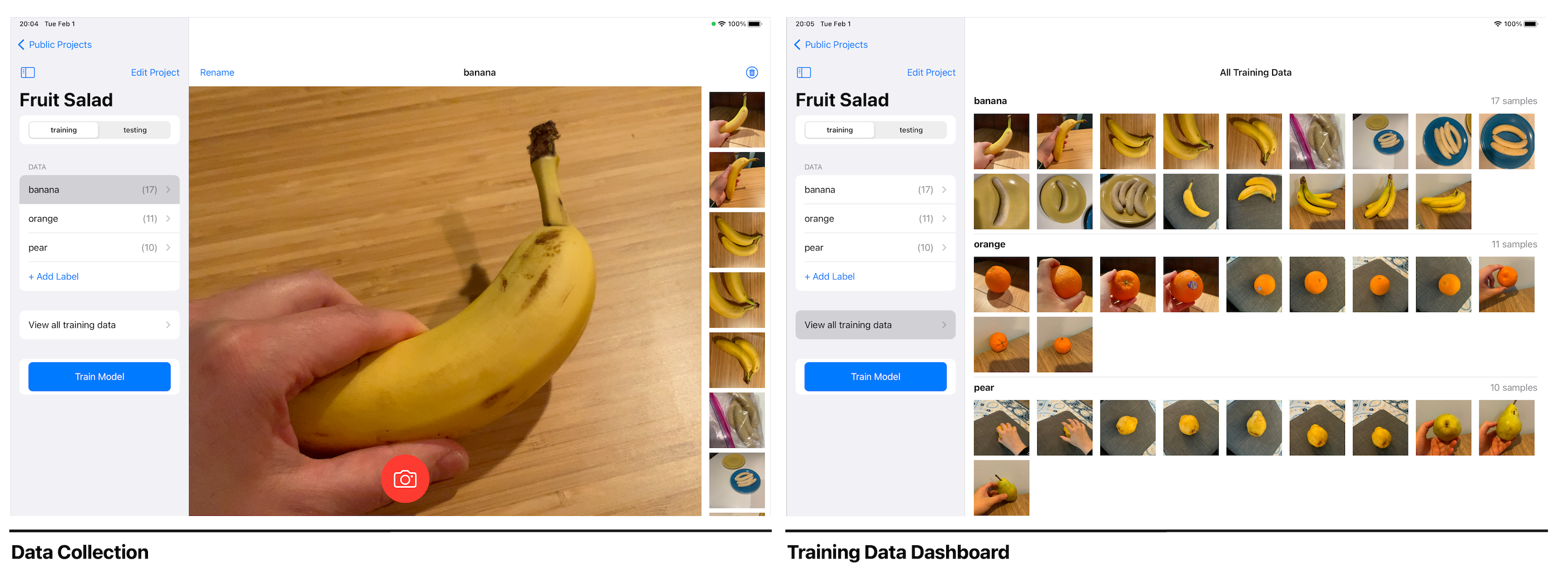}
  \caption{Data collection interface for adding labelled images to the shared dataset (left).  All images added across devices are visible in the synchronized Training Data Dashboard, organized by label name (right).}
  \Description{The Data collection interface and training data dashboard for Fruit Salad project}
  \label{fig:training-mode}
\end{figure*}

\subsection{Collecting and Reviewing Training Data}
Tapping a label name opens a camera interface for collecting labelled photographs (Figure \ref{fig:training-mode}). Existing data for a given label are displayed alongside the camera to support users adding new images. The relative sample count for each label is displayed alongside the label name to enable quick inspection of class balance. Users can tap an image and remove it from the dataset if they chose to. Tapping the \textbf{View all training data} button displays all training images in a grid of thumbnails grouped by label name as shown in Figure \ref{fig:training-mode}. Collaborative data collection is supported by synchronizing data across all devices and presenting data in a single dashboard, inviting users to review and identify patterns and gaps in the dataset. Image data are stored and synced using the CloudKit service as the back-end \cite{cloudkit}, with images stored in private CloudKit data stores accessible only to the users within a project. 

\subsection{Training a Model}
Tapping the \textbf{Train Model} button generates an ML model using transfer learning and the Create ML Image Classifier API, a Swift framework for training Core ML models \cite{createml, coreml}. The model is then consumed using the Core ML API to classify new data. On-device training typically takes about 5 seconds for training data consisting of a several hundred images using 2019 or newer iPads.

At the moment, Co-ML centers the ML model building process around working with data (its collection, cleaning, and analysis), so we do not expose the underlying model architectures and algorithms for a trained model. As it is often easier to improve model performance by iterating on data rather than tweaking model architecture or algorithms more broadly \cite{hohman2020understanding}, we believe that focusing user exploration on dataset design offers a powerful entry point for beginners to start to reason about how ML models work.

\subsection{Testing the Model}
\begin{figure*}
  \centering
  \includegraphics[width=0.9\linewidth]{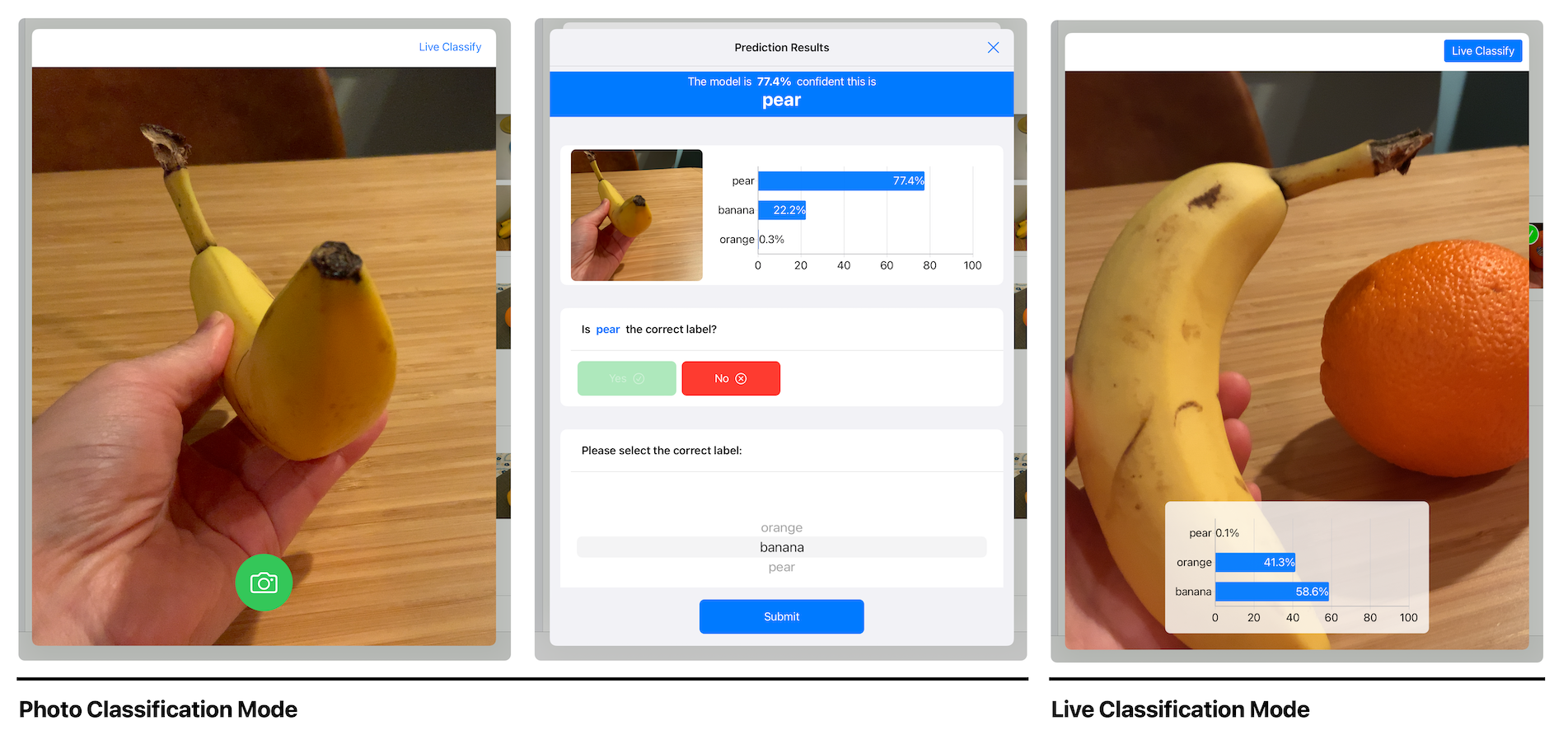}
  \caption{Classifying new data using the Photo Classification Mode (left) and Live Classification Mode (right). In Photo Classification Mode, the user takes a photograph and can review the classification results. In Live Classification, users can see an updating bar chart displaying relative confidence levels for each class.}
  \Description{\todo{add description}}
  \label{fig:classification-mode}
\end{figure*}

After a model is trained, the app presents the user with a camera interface for classifying new data as shown in Figure \ref{fig:classification-mode}. Users can test the model's performance in two ways: photo mode and live classification. In the default photo mode, the user captures an image and is presented with a summary of classification results, including a bar chart of relative confidence levels for each label. If the model is wrong, users can provide the correct label. All images taken in this mode are added as test data to assess model performance across iterations (described in the next section). A secondary Live Classification mode enables users to classify data in real time and test boundary conditions where the model transitions from one classification to another. We chose to default to the photo classification mode rather than live classification (the default mode in existing tools like Teachable Machine \cite{carney2020teachable}) because 1) we want to support beginners actively forming hypotheses before seeing the classifier results, and 2) users can review and discuss results without having to interpret a constantly updating confidence barchart.

\begin{figure*}
  \centering
  \includegraphics[width=\linewidth]{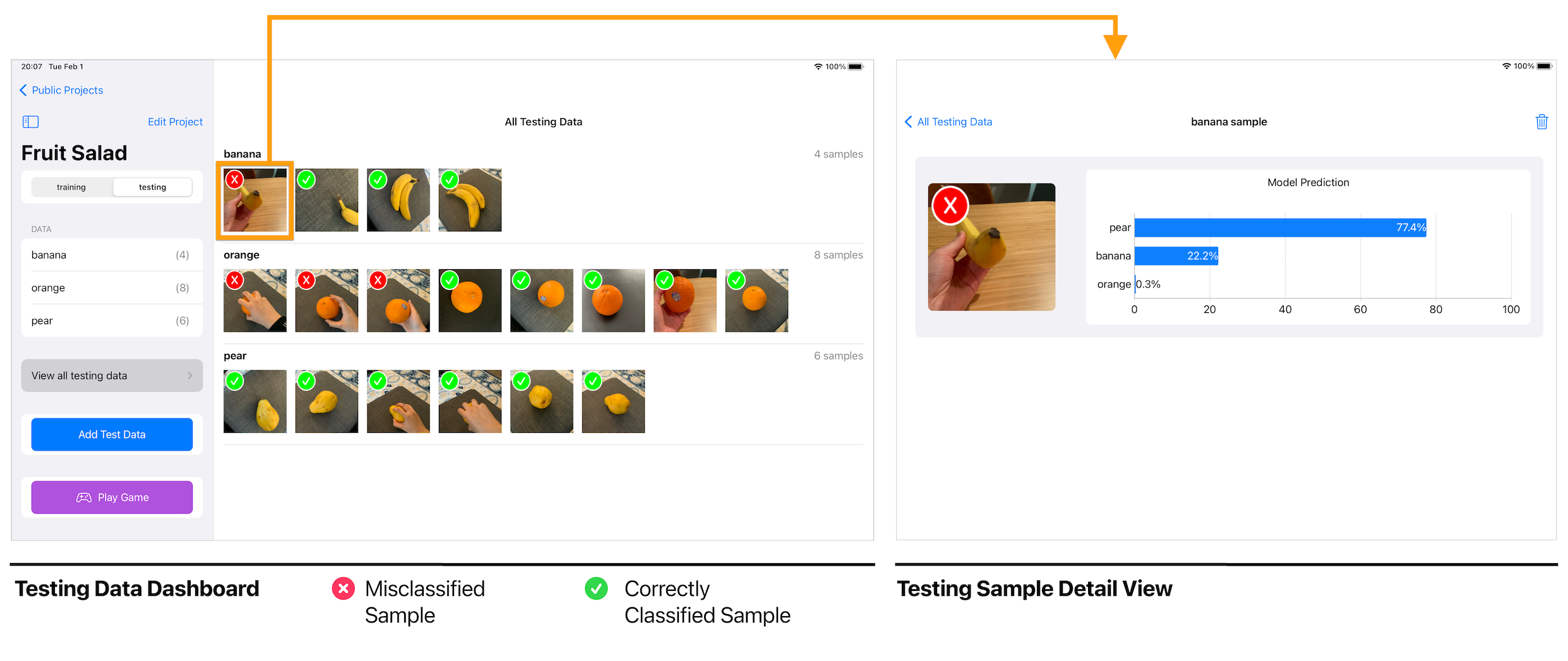}
  \caption{Testing mode interfaces.  Users can review collectively added test data, and classification results are based on the latest trained local model. Tapping on a misclassified sample shows a bar chart of confidence levels to help users debug model performance.}
  \Description{\todo{add description}}
  \label{fig:testing-mode}
\end{figure*}

Closing the classification interface presents the testing dashboard, where test images collected during photo classification can be reviewed (Figure \ref{fig:testing-mode}). Similar to training data, we support collaboration by synchronizing test data to encourage all users to share and reflect on their collective model testing results. Each test image thumbnail has a badge indicating if the model classified the image correctly or not (with a green checkmark or red X), with misclassified images grouped first. Tapping a test image presents a barchart of classification results. Whenever the model is retrained, all test images are automatically re-classified using the latest model, and their corresponding badges are updated; users can thus see whether previously misclassified images are correctly classified after iterating on their model.

\subsection{Evaluating the Model: Restaurant Frenzy Game}

To support playful model testing, we created an in-app game called \textit{Restaurant Frenzy}, inspired by the popular games Diner Dash \cite{diner-dash} and Overcooked \cite{overcooked}. In the game, the player is a cook trying to complete as many orders as they can in a fixed time period (90 seconds) by showing the camera the needed ingredient. The 90 second time limit supports testing each label several times. We designed a food-themed game because cooking is often a shared experience for families, it can use inexpensive and easily obtainable ingredients, and physical objects naturally introduce variables important for quality image datasets like perspective, lighting, and size.

\begin{figure*}
  \centering
  \includegraphics[width=\linewidth]{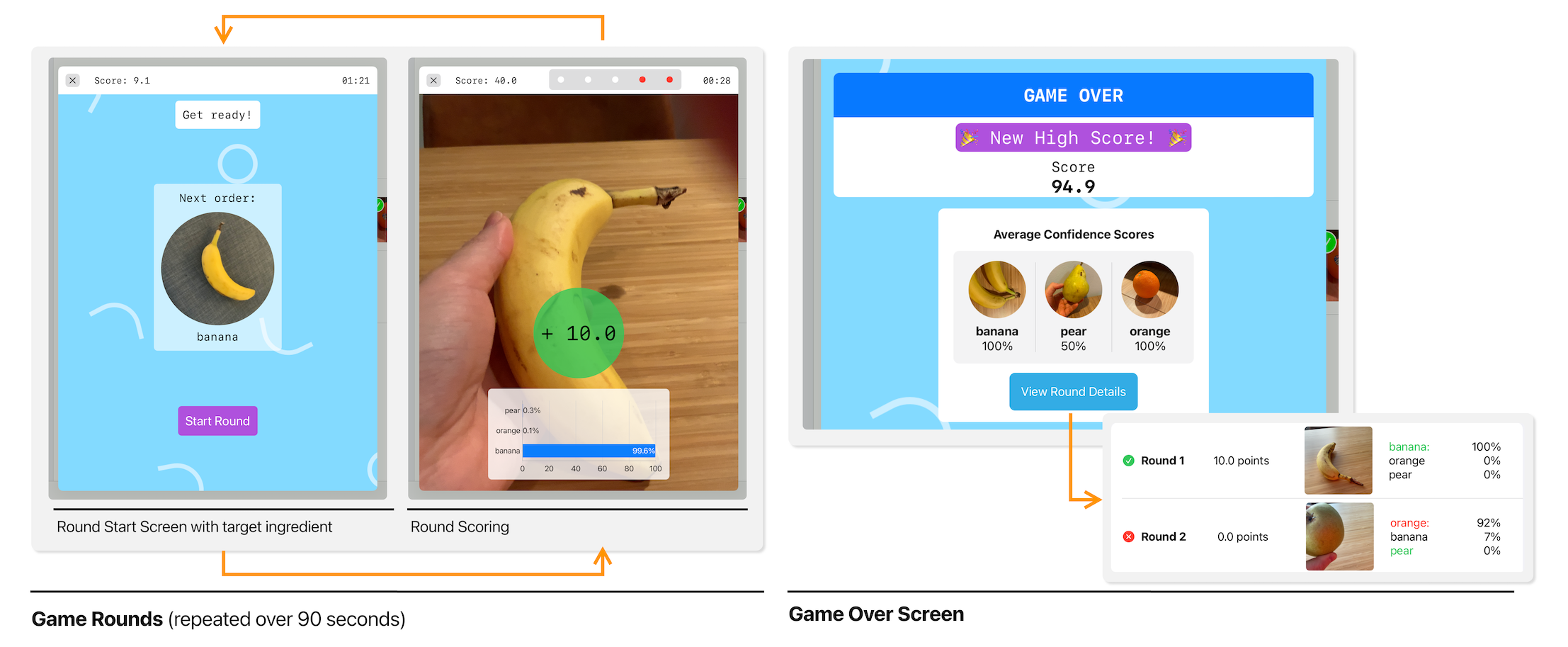}
  \caption{Restaurant Frenzy Game. Users fulfill as many rounds as they can within a 90 second time limit. In each 5 second round, the user presents a target object to the camera and is scored by the confidence level of the classification. After the game is complete, a Game Over screen displays their total score and average confidence levels per label. Tapping the View Round Details button provides information about model performance in each round of the game.}
  \Description{\todo{add description}}
  \label{fig:game-ui}
\end{figure*}

Every round presents players with a random target label, and they have 5 seconds to present the object to the camera (Figure \ref{fig:game-ui}). During each round, a barchart of live classification results is displayed alongside the camera, and round scores are calculated using the model's confidence at the end of the round (a confidence level of 78\% gets 7.8 points, and a user scores 0 points if the model misclassifies the object). At the end of the game, a Game Over screen shows their final score and the average confidence per label. Tapping the \textbf{View Round Details} button lets users review specific rounds of the game to see when items were misclassified. The game is designed to encourage collaborative play (with one person controlling the iPad, and others positioning items for the camera to classify) and collective discussion to identify failure cases, which families can then use to decide how to improve their model.  

\subsection{Iterating on the Model}
Users can revise their model by adding or removing training data, retraining their model, and seeing whether the model improved. To foster collective ownership of the dataset, all users have equal access to editing the project including adding new labels, renaming or deleting existing labels, and adding or removing image data. Improvement can be seen by reviewing whether test data classifications results have changed after re-training or by re-playing the game and seeing if they can beat their high score.
\section{Methodology}
\label{sec:methodology}

We recruited families with children ages 10-14 to try out Co-ML and our companion activity. Families were recruited via a database of employees at Apple who had previously expressed interest in participating in research studies. We used a recruitment survey to screen for a number of factors including: ethnicity, number of parents/guardians, number of children and their ages and gender identities, and the profession(s) of the parent(s) and their prior exposure to ML. We chose families with two children and at least one parent who could participate. Once families were selected, we lent them 11'' iPads with the Co-ML app pre-installed, providing a tablet for each participating family member. Each family participated in a single 2-hour session facilitated via video conference to maintain safety during the COVID-19 pandemic.. Each participant was compensated with a \$40 gift card.  

\subsection{Session Structure and Activity}
In advance, we asked each family to decide on a favorite family dish and gather 5-8 items (ingredients, utensils, and spices) to use during the activity. We conducted sessions remotely through video-conferencing using the same activity script for each 2-hour session. Three researchers were present: One researcher facilitated the session and engaged with the participants, the second took observation notes, and the third provided technical support. All sessions were conducted in English.

At the start of each session, we introduced the research team, answered any questions about the consent and assent forms, and ensured the forms were reviewed and signed by each participant before proceeding. After the family shared the family dish they selected (e.g., pizza) and what ingredients they had (e.g., pita bread, tomato sauce, and mushrooms), we presented a brief demo of the Co-ML app before asking participants to narrow their pool of collected ingredients (and thus, the labels for their model) to 3-4 items (one per family member). Next, each participant received their own iPad and were instructed to take 5 photos of one item using their own device to add to the training dataset. The facilitator then guided a discussion in which participants reviewed their collective dataset. Next, we asked family members to swap items and take additional photos until each label had at least 15 images. Families reviewed the images again as a group and were asked if they could anticipate any sources of confusion for the model based on the data. We then guided the family to train their model and explained how to take and review testing data. Families tested every item in their classifier and discussed whether the model was correct or incorrect in its predictions.

The family then used a single iPad to play the Restaurant Frenzy game to evaluate their model, and after playing the game once, families were given 15 minutes to freely iterate and try to improve their models together. They then played the game again to see if their revised model resulted in a higher game score. At the end of the session, families completed a semi-structured interview about their experience.

\subsection{Data Collection and Analysis}

After confirming consent to do so, we video-recorded each session using the video-conferencing platform and collected screen recordings of each individual's iPad to understand how each participant iterated on their models. We supplemented our video data with logs of timestamped records on CloudKit showing when individuals added or removed data. Data families added to Co-ML were only viewable to other family members and our research team. Upon completion of all sessions, we exported image data from CloudKit, encrypted the photos so they were only available to the research team, and deleted all cloud-stored data for participants' privacy.

Four researchers analyzed the data, which involved 1) stitching together the video recording from the video conference with individual iPad screen recordings to analyze each participant's actions during the activity; 2) transcribing the conversations and deductively coded for ML model building concepts that emerged when users verbally discussed strategies for improving their model; 3) compiling timestamped events from the stitched video along with CloudKit event records to create visualizations of how users navigated through Co-ML in support of their model building process. Researchers met for periodic analysis discussions to reach consensus around our interpretations, including families' interactions with each other and with Co-ML.

\subsection{Case Study Selection}
\label{sec:methodology-case-study}
As an exploratory study, our aim is to illustrate the feasibility and potential benefits of collaborative ML modeling with Co-ML through analyzing learners' interactions while building ML models together. Because we are concerned with understanding how and why individuals refined their approaches to dataset design throughout the activity, we chose a case study approach \cite{yin2009case} that could support rich description of the multi-user modeling strategies observed. 

 While we analyzed sessions of all families that participated, we largely focus our description in the remainder of the paper on a single family that uniquely met the following criteria: 1) each family member actively spoke during the activity, allowing us to analyze how they reasoned about ML through their discourse with one another; and 2) each family member used their own iPad continuously throughout the activity, enabling a finer-grained analysis of individual-device log data to determine how each family member contributed to model building. Centering on a single family enables us to most clearly illustrate how individual ideas around building balanced datasets changed over the course of iterating on a model together. 
 
 We follow our in-depth case study with a brief summary of the remaining families for context; however, an in-depth comparison is out of scope for this work. Examples of single case studies previously published in the IDC community include \cite{bell2015learning, davis2017haptic, di2009interactive, lyons2015designing, yip2014helped}. Through this proof of concept, we lay the groundwork for future research about how dataset design practices can be facilitated through multi-user activities with technologies like Co-ML; we plan to further investigate how different types of activity structures and group compositions might impact learning of ML in future work.

\section{Results}
\label{sec:results}
Three families (6 children and 5 parents) participated in our Co-ML activity, building food classifiers for pizza, salad, and spaghetti ingredients.  Over the course of playing the Restaurant Frenzy game twice, families created training datasets ranging from 126-182 total images, with 2/3 families increasing their score in the game after iterating on their datasets.

For clarity of discussion, we begin by presenting an in-depth case study (\ref{sec:methodology-case-study}) that illustrates how the Spaghetti family learned about dataset design (including diversity, representation, and class balance) through a collaborative model building activity using Co-ML.  As a single case study, we do not intend to generalize these findings to all families, but rather to provide rich description illustrating how using and discussing Co-ML helped surface different perspectives and opinions on data strategies. Our goal is to support active discussion between family members to enable multiple perspectives to be represented in the data; a case study analysis of this specific family's successful interactions can support scaling the work to all families in the future. Following discussion of the case study, we briefly summarize our observations of the other two families (Pizza and Salad) that participated.

Our case study involves a family of four: YS (Younger Son, 11 years old), OS (Older Son, 14 years old), Mom, and Dad. They identified as Latino and White, and Mom and Dad work professionally as a product designer and software engineer respectively. For their favorite dish spaghetti, the family chose sauce, spaghetti pasta, pot, and spoon as labels for their classifier. Through this case study, we highlight this family's collective actions to resolve a ``spaghetti misclassification,'' which in turn surfaced multiple ML dataset design strategies.

\subsection{Initial Training Data Collection} 
\label{sec:results-initial-training}
\begin{figure*}
  \centering
  \includegraphics[width=\linewidth]{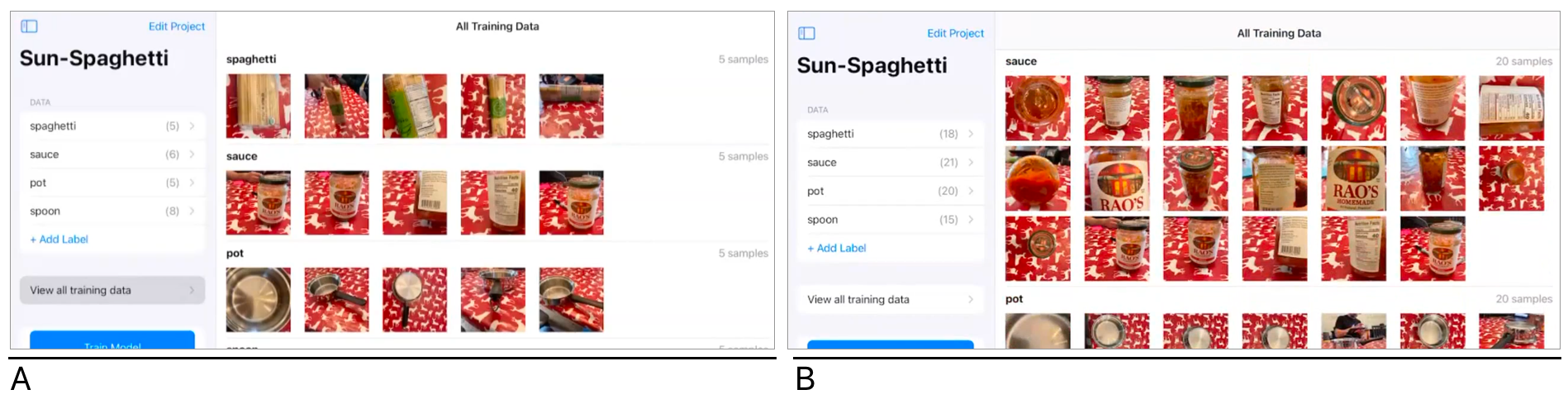}
  \caption{A. Initial training data before discussion. B. Expanded training data for Sauce added after discussion, where photos of the top and front-labels of the jar were added.}
  \Description{}
  \label{fig:moment-1}
\end{figure*}

At the start of the activity, each family member chose an ingredient and added an initial set of 5 training images (Figure \ref{fig:moment-1}A). The researcher then asked them to review the images together: ``You're going to be adding a few more images before we train the model.  Given the images here, do you think there are any other helpful images that you would want to include in your training dataset?'' In response, Dad offered that images of the top and bottom of sauce jar could be helpful, but Mom disagreed: ``Unless it matters the brand or the type, that would be what the top of the jar would give us. But if it's just that we're trying to confirm it's the sauce, I don't think it's necessary.'' At the same time, OS shared that Mom had taken three pictures from the same side (the front label of the jar), perhaps suggesting that different looking images may be beneficial. Their discussion of \textit{features} of the image data that may be important for \textit{representation} of the sauce shows how each family member was hypothesizing about what may be more or less helpful for the model.

Next, we asked family members to swap ingredients and add 10 more images of their new item. YS was given the jar of sauce and added images that included the top and bottom of the jar, indicating that he had taken on board suggestions from the family's discussion on how to expand their dataset (Figure \ref{fig:moment-1}B).  During this process, OS continued to show his careful attention to the images by reporting that spaghetti images had been incorrectly added to the spoon label and asking who would retake the images. YS offered to retake the spaghetti images, Dad provided advice on how to prevent the mistake from happening again (by pointing out where in the UI the label name is shown), and Mom invited others to proactively delete mislabeled images directly.    

\begin{figure*}
  \centering
  \includegraphics[width=\linewidth]{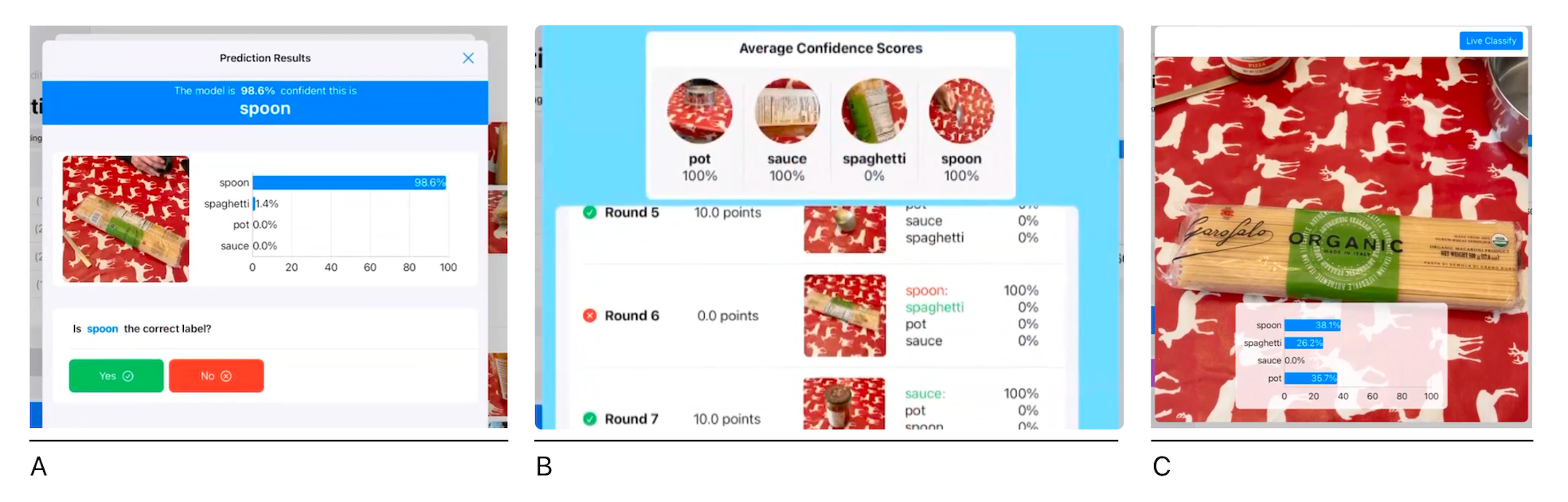}
  \caption{A. The first misclassification of spaghetti as spoon. B. The game over screen showing misclassified rounds. C. Using live classification to find where the model switches between spaghetti and spoon classifications.}
  \Description{}
  \label{fig:spaghetti-moments}
\end{figure*}

\subsection{First Misclassification and Resulting Debugging}
\label{sec:results-first-misclassification}
The facilitator asked the family to train their model and test it in Photo Classification mode. The spoon, sauce, and pot were each correctly classified with high confidence, but spaghetti was misclassified as spoon as displayed in Figure \ref{fig:spaghetti-moments}A. This misclassification set off a series of model testing trials in which several theories emerged. Mom shared that when a new item is introduced, the camera refocuses on the newer object before classifying, while YS agreed by added that the camera may become blurry in the process of refocusing.  Dad described how the ``shininess of the spoon is grabbing the camera's attention more than spaghetti.'' OS offered that the similarly long shape of the spoon and spaghetti may an issue. Their discussion surfaced competing theories ranging from similar properties of the object (their reflective qualities and shape), properties of the image (whether it is blurry or not), and properties of the sensor (camera) classifying the object (whether classification depends on what the camera focuses on). 

\subsection{Game Play 1}
\label{sec:results-game-play-1}
When the family played the game, they found that spaghetti was misclassified again (all other items were classified correctly with high confidence). After reviewing the game over screen (Figure \ref{fig:spaghetti-moments}B), they began to discuss how multiple properties, or data \textit{features}, in combination may play a role; when YS brought up that the pot is also shiny like the spoon, Dad said that the length of the spaghetti and spoon were both long, suggesting that since the pot was not similarly long, it would not be confused for spoon. Mom then described a realization: ``When we were doing the exercise [data collection], I said I don't think it matters if we take a picture of the top [of the sauce jar], but watching Dad in the game, everything was from the top down. So now I actually think taking a top-down picture of the jar was more important than I thought originally.''  Through seeing how the model was used, she was able to revise her original ideas about what types of data variation are valuable.  

\subsection{Live Classification: Spaghetti Zoom Insight}
\label{sec:results-live-classification}
Next, the researcher introduced the family to the Live Classification to view classification results in real-time.  They proceeded to test this feature with the spaghetti and spoon as shown in Figure \ref{fig:spaghetti-moments}C, where Dad moved the iPad closer and farther away from the packaging and witnessed a boundary condition where the classification toggled between spoon and spaghetti. When the researcher asked for ideas for when the classification switches, YS replied, ``When it sees the entire thing [spaghetti packaging],'' sharing his insight that the misclassification only occurs when more of the spaghetti packaging is visible.  Mom then offered her own debugging suggestion: ``Does it do the same thing with the spoon, or is it just confused by the spaghetti?''  Dad and YS respond that it only misclassifies with the spaghetti, and Dad showed how zooming in and out of the spoon does not change the model's classification of the spoon.  Through their discussion, they decided that the spaghetti data needed to be improved more than the spoon data.

\subsection{Dataset Gap Realization}

\begin{figure*}
  \centering
  \includegraphics[width=\linewidth]{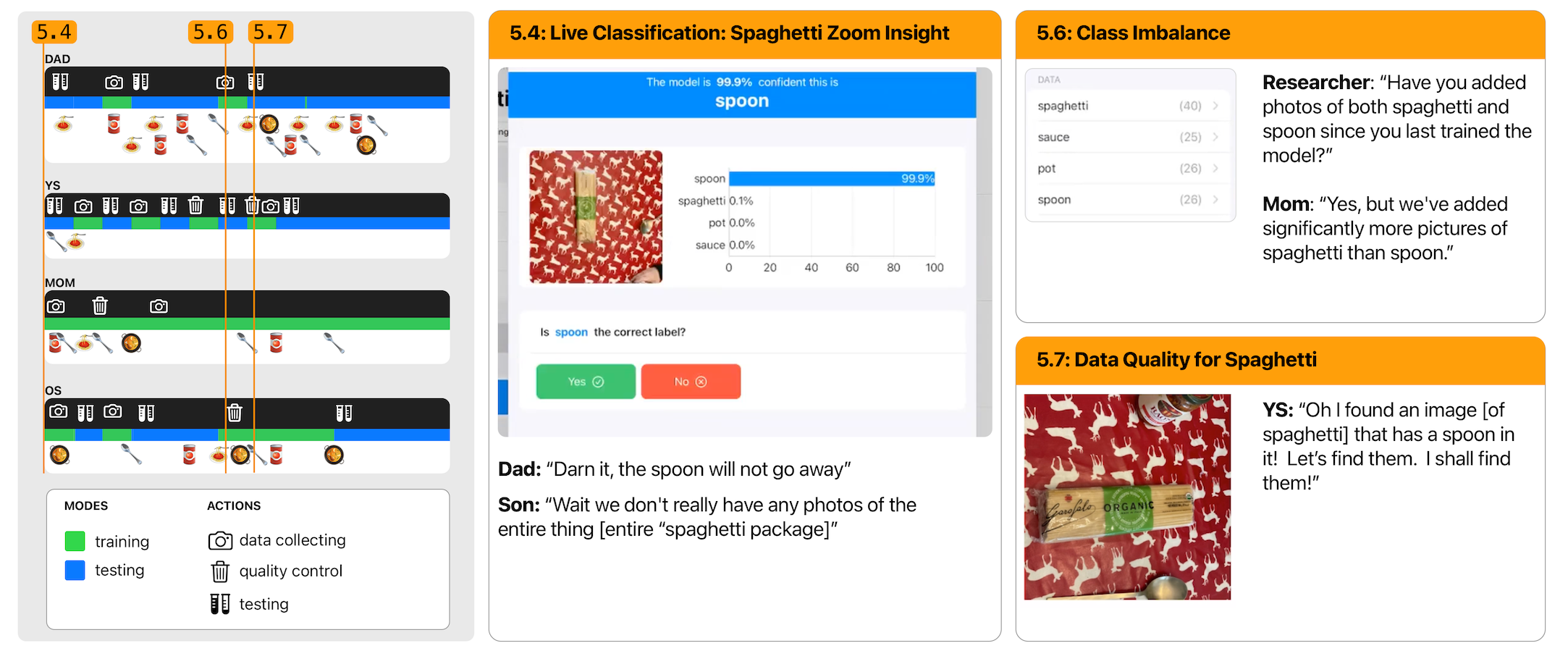}
  \caption{Family A iteration moments during 15 minutes of free play, displaying how individual family members chose to iterate on the model through data collecting, quality control, and testing. The ingredients the family members chose to work with (spaghetti, sauce, pot, and spoon) are also displayed. Three moments are highlighted, alongside screenshots from the Co-ML app, that show how the family reasoned about their model's performance.}
  \Description{A figure displaying, for each family member, which model collecting action they took on throughout the activity, as well as which labels they attended to and which mode they were in (training or testing). \todo{update moment numbers}}
  \label{fig:family-a-swimlane}
\end{figure*}

Over the next 15 minutes, the family iterated on their model however they chose to do so. Unlike the rest of the activity up to this point, this period was unfacilitated and thus provided a means to observe choices individual family members took to improve the model, as well as how they worked together to address their spaghetti misclassification. In Figure \ref{fig:family-a-swimlane}, we represent the family's parallel efforts during this free play period, categorizing three different types of modeling actions: \textbf{Data Collecting} - Adding new training data; \textbf{Model Testing} - Using classification to identify where the model fails; \textbf{Quality Control} - Culling through the data dashboard and selectively removing images to clean the data. Figure \ref{fig:family-a-swimlane} also shows which items each family member was attending to over the 15 minutes as well as how often they switched between training and testing, with all but Mom shifting between the two throughout this period. For context, we provide three markers indicating moments during this period that we describe in the following sections.

Dad added training samples of spaghetti, while YS identified that the model classified an image of the family's vivid tablecloth (on which all other training samples were taken), without any items present, as spoon.  After YS reviewed all the training data for spaghetti, he noticed, ``We don't really have a lot of photos of the entire thing [spaghetti packaging],'' and pointed out that the majority of the existing training images consist of zoomed in details on the spaghetti packaging. YS then went on to take 23 images of the spaghetti from above, capturing the entire package in each photo: ``I'm going to figure out this mystery.  I'm trying to become George...Curious George!'' At the same time, Mom continued to test the model's performance on both spoon and spaghetti. In doing so, the two members are each tackling the same issue from multiple angles: data collection and testing.

\subsection{Class Imbalance}
\label{sec:results-class-imbalance}
At this point, YS had added more images of spaghetti (there were 41 training images for spaghetti but only 23 for spoon). As the family continued debugging, the researcher asked, ``So have you added photos of both spaghetti and spoon since you last trained the model?''  In response, Mom said, ``Yes, but we've added significantly more pictures of spaghetti than spoon,'' and YS agreed: ``A lot!'' This realization about \textit{class imbalance} (that there were more spaghetti images than spoon) kicked off further data collection from Dad, who captured additional training data for spoon.

\subsection{Data Quality for Spaghetti}
\label{sec:results-data-quality}
YS, who continued reviewing spaghetti data with the training data dashboard, then declared, ``I found a [spaghetti] image with a spoon in it!'' He showed Mom an image where the spoon was on the edge of a spaghetti sample, and then deleted it. \textit{Quality control} actions were then taken on by both YS and OS, where both actively cleaned the dataset by removing misrepresentative images. While OS was relatively quiet throughout the activity, our screen recordings revealed his active participation, where he took on the most frequent number of Quality Control actions. Collectively, OS deleted a total of 28 images, identifying samples where multiple items were visible (for spaghetti as well as for spoon and sauce). As they continued over the next four minutes, YS shared that after retraining, the model had improved.

\subsection{Game Play 2 \& Activity Reflection}
\label{sec:results-game-play-2}
At the end of their iteration period, the family had increased their training dataset from 74 total images to 126 by the time they played Restaurant Frenzy again. The first round spaghetti was the target ingredient, YS stated, ``The hard one...this is the moment of truth!'', and when the classifier returned a confidence of 100\% for spaghetti, the family cheered. But in later rounds, the model was not as successful – spaghetti was misclassified as spoon in another round, and pot was also misclassified as spoon. While the family got a slightly lower score in their second play of the game, the average confidence level for spaghetti improved from 0\% in the first gameplay to 68\%, suggesting they had improved the model's performance on spaghetti overall. 

When the family reflected on their experience at the end of the session, the researcher asked, ``Do you think adding more data affects the accuracy of the model?'' In response, YS shared, ``The extra data could also change other data...because if you fix one thing, the thing that it takes to fix that could cause another problem with another thing.'' Here, the child revealed an understanding that simply adding more data to a model does not necessarily improve how well the model performs. Improving performance for one label can cause regressions for other labels \cite{sculley2015hidden}, a critical insight that the 11-year-old was able to develop in his limited experience using Co-ML.

In their discussion, YS shared a breakthrough in his understanding of the spoon's role in misclassification: ``Because the spoon is the thinnest thing of all, the reason it [the model] thought the other things were the spoon is that the more we zoomed out, the more it saw the background. Since the spoon is the smallest thing, it shows more of the background.'' This related to his insight earlier where a photo of the tablecloth classified as spoon, identifying a spurious relationship in which the model is classifying the tablecloth as spoon. Mom added that it would be important to teach the computer ``how to ignore the busy background.''

Mom described how Co-ML supported each son's preferred approach; while OS was less talkative than his brother, he contributed in his own way by, ``catching people putting pictures in the wrong category and combing through the data and deleting data points that he thought were not quality for the exercise.'' Through YS's continual discussion with his parents and OS's more quiet data cleaning practices, the family collectively debugged their model.

\subsection{Comparison to Other Families}
While presenting our full analysis of all three families is out of scope for this work, here we summarize some of the main findings for the Salad and Pizza families for context.

Similar to the Spaghetti family, the Salad and Pizza families also focused on data representation, noting the unique packaging of their ingredients (e.g., cans, jars, bags or printed labels). Further, Salad family engaged with the idea of class imbalance. After noticing that they had significantly more samples for Spoon than their other labels, the son wondered whether this might hurt the model's performance: ``If you took hundreds of pictures of the spoon, that means basically [if] any part of the spoon [is] showing, it [the model] would instantly recognize it. So if you tried adding a different thing that maybe had a similar color and also reflected, it would think that it’s a spoon. In some ways, it would hurt it [the model] instead of make it better.'' Unlike the Spaghetti family, who mainly took photos of their objects placed on their dining table, the other two families ended up utilizing diverse backgrounds either unintentionally (due to where people were positioned, such as working on different surfaces or against different backdrops) or through intentional design (at one point the Salad family tested their model by holding objects up against a bright yellow hoodie to test whether the color of the background mattered). Both Spaghetti and Salad families actively engaged in discussion with one another, with individuals fluidly switching between training and testing and working with multiple ingredients. In contrast, the Pizza family preferred to work silently, with each person focused on debugging one single ingredient for the entirety of the free play period. We wonder whether promoting collaborative dialogue and interactions may help encourage engagement with multiple stages of the model-building process. Additional studies and further work would be needed to explore and verify this hypothesis.

\label{sec:results}
\section{Discussion}
\label{sec:discussion}
The multi-user modeling experience of Co-ML supported the collaborative collection, review, and revision of data, enabling the family to encounter core dataset design considerations of data representation, diversity, and class balance. Notably, these dataset practices are foundational for addressing existing biases in AI systems as discussed in Section \ref{sec:related-work-ai-literacy}, so we are encouraged that this short activity using everyday objects could facilitate learning about these considerations.

\textbf{Data Representation and Diversity.} Since the family had one item for each label, their \textit{dataset diversity} included multiple camera angles (top, side, or bottom views for objects) and different zoom levels (close up pictures versus showing a whole object). Discussion over the synchronized dataset dashboard in Co-ML helped family members notice and consider types of data representation. Attributes like color, length, and shininess were discussed as \textit{features} the model might use to tell items apart. These discussions helped surface missing, erroneous, or underrepresented data such as top-down images of the sauce jar or mislabeled images (\ref{sec:results-initial-training}). Dataset gaps were further identified through discussion about the Game Over Screen to see where the model had the poorest performance (\ref{sec:results-game-play-1}) and using the Live Classification feature to realize that the dataset was missing images of the entire spaghetti packaging (\ref{sec:results-live-classification}). These results illustrate how a collaborative experience, specifically a shared dataset and collective discussion, can help surface differences in data collection approaches that can ultimately inform and enhance ML datasets. 

\textbf{Class Imbalance}. Along with considering diversity within a single label, the family also recognized that balancing datasets \textit{across} labels was important. Mom and YS recognized the class imbalance of having far more images of spaghetti collected than spoon, which led Dad to capture more images of the spoon (\ref{sec:results-class-imbalance}). Yet, it was through direct experimentation that YS identified that adding more data can cascade into other issues (\ref{sec:results-game-play-2}), going beyond the common misconception that ML models can be improved just by adding more data \cite{yang2018grounding}:``Because the extra data can change other data. If you fix one thing, the thing it takes to fix that could cause another problem with another thing [label]'' (YS). This is a common problem that professional ML engineers wrestle with –– that adding data alone does not necessarily improve a model because of the introduction of new issues \cite{hohman2020understanding}. 

\textbf{Data Quality.} When YS and OS each identified issues with data quality (mislabeled images as in \ref{sec:results-initial-training} or images with multiple objects visible as in \ref{sec:results-data-quality}), YS, OS, and Mom contributed to cleaning the dataset while Dad helpfully point out where to confirm images were added to the correct label in the Co-ML interface. Most deleted images were taken by another family members (rather than the user taking on the data quality role), supporting our idea that having multiple people contribute to the data helps learners encounter viewpoints different than their own.

\textbf{Overfitting.} Overfitting, or situations where the model learns patterns in the training data so closely as to not generalize to new contexts, did not arise in the family’s conversations, but ended up being a challenge with the study activity more broadly. Because the family used the same items for both training and testing their model, and trained and tested their model in the same place (around their kitchen table), the model they developed likely overfit to a narrow set of use cases. Overfitting could be potentially reduced if, 1) families were asked to reserve a similar but distinct version of an item to use for testing (such as having 2 jars of sauce from two different brands, with one only used for testing), and/or 2) the family was asked to test with a different background (such as a different countertop in their house). Future work could adapt this activity to introduce novel contexts to reduce overfitting and support testing how well a model generalizes.

\textbf{Multiple modes of participation}. We observed that Co-ML supported equitable and differentiated approaches to model building. Since no single user has sole control or influence over the developing dataset, each user can decide which ideas and priorities they themselves want to pursue, encouraging different styles of participation. In summarizing the distinct preferences of OS and YS, Mom said, ``I think a good distinction is our two kids.  We had one YS who was very interactive, and one OS that was not verbally interactive but was really more focused on data quality...focused on catching people putting pictures in the wrong category and combing through the data and deleting data points that he thought were not quality for the exercise.'' Multiple ways of participating suggests that even for those who are not as involved in the verbal discourse, listening to other people's concerns can still enable individuals to act upon new ideas.
\section{Conclusion}
We presented Co-ML, a novel tablet-based application for beginners to build and test ML image classifiers using a collaboratively built dataset. Through an in-depth case study, we presented how a family using Co-ML encountered critical ideas in ML dataset design, including data representation, diversity, and class imbalance. These design considerations were fundamentally shaped by having multiple points of view from each of the family members represented in the data, enabling different strategies for training and testing models to be debated and explored to ultimately diversify their dataset. Further, through a distributed modeling experience with Co-ML, family members were able to work in parallel through multiple stages of the model-building pipeline, including data collection, testing, and quality control, showing that these fundamental ML modeling practices can be supported with appropriately designed tools and activities. Through this work, we contribute a hands-on, collaborative approach to introducing strategies for creating balanced datasets and ultimately hope to encourage more people to consider collaborative, social learning as a powerful means to broaden ML participation and literacy.

\label{sec:conclusion}

\section{Selection and Participation of Children}

Families were recruited via a database of employees at Apple that had expressed interest in participating in research studies.  We selected families after they completed an initial recruitment survey with questions such as their ethnicity, number of children as well as their ages and their gender identities, and the professions of their parents.  At the beginning of our virtual sessions, a researcher went over consent forms and answered any questions from the family, and children and their families were each required to sign individual consent forms before we proceeded with the study. The protocol and data collection for this research study was reviewed and approved by a research ethics committee and legal counsel at a large technology company.  All data collected using the Co-ML app was only visible to members of the family and the research team, and upon completion of the study, data was removed from Cloud-based data stores, encrypted, and made available only to our research team for analysis.

% \begin{acks}
% \todo{We thank our colleagues at Apple for their time and effort...}
% \end{acks}

\bibliographystyle{ACM-Reference-Format}
\bibliography{co-ml}


\begin{thebibliography}{48}

%%% the Web address.

\ifx \showCODEN    \undefined \def \showCODEN     #1{\unskip}     \fi
\ifx \showDOI      \undefined \def \showDOI       #1{#1}\fi
\ifx \showISBNx    \undefined \def \showISBNx     #1{\unskip}     \fi
\ifx \showISBNxiii \undefined \def \showISBNxiii  #1{\unskip}     \fi
\ifx \showISSN     \undefined \def \showISSN      #1{\unskip}     \fi
\ifx \showLCCN     \undefined \def \showLCCN      #1{\unskip}     \fi
\ifx \shownote     \undefined \def \shownote      #1{#1}          \fi
\ifx \showarticletitle \undefined \def \showarticletitle #1{#1}   \fi
\ifx \showURL      \undefined \def \showURL       {\relax}        \fi
% invisible to TeX
\providecommand\bibfield[2]{#2}
\providecommand\bibinfo[2]{#2}
\providecommand\natexlab[1]{#1}
\providecommand\showeprint[2][]{arXiv:#2}

\bibitem[Agassi et~al\mbox{.}(2019)]%
        {agassi2019scratch}
\bibfield{author}{\bibinfo{person}{Adam Agassi}, \bibinfo{person}{Hadas Erel},
  \bibinfo{person}{Iddo~Yehoshua Wald}, {and} \bibinfo{person}{Oren
  Zuckerman}.} \bibinfo{year}{2019}\natexlab{}.
\newblock \showarticletitle{Scratch nodes ML: A playful system for children to
  create gesture recognition classifiers}. In
  \bibinfo{booktitle}{\emph{Extended Abstracts of the 2019 CHI Conference on
  Human Factors in Computing Systems}}. \bibinfo{pages}{1--6}.
\newblock


\bibitem[Apple(2023a)]%
        {cloudkit}
\bibfield{author}{\bibinfo{person}{Apple}.} \bibinfo{year}{2023}\natexlab{a}.
\newblock \bibinfo{title}{CloudKit}.
\newblock
\newblock
\urldef\tempurl%
\url{https://developer.apple.com/icloud/cloudkit/}
\showURL{%
\tempurl}


\bibitem[Apple(2023b)]%
        {coreml}
\bibfield{author}{\bibinfo{person}{Apple}.} \bibinfo{year}{2023}\natexlab{b}.
\newblock \bibinfo{title}{CoreML}.
\newblock
\newblock
\urldef\tempurl%
\url{https://developer.apple.com/documentation/coreml}
\showURL{%
\tempurl}


\bibitem[Apple(2023c)]%
        {createml}
\bibfield{author}{\bibinfo{person}{Apple}.} \bibinfo{year}{2023}\natexlab{c}.
\newblock \bibinfo{title}{CreateML}.
\newblock
\newblock
\urldef\tempurl%
\url{https://developer.apple.com/machine-learning/create-ml/}
\showURL{%
\tempurl}


\bibitem[Barbu et~al\mbox{.}(2019)]%
        {barbu2019objectnet}
\bibfield{author}{\bibinfo{person}{Andrei Barbu}, \bibinfo{person}{David Mayo},
  \bibinfo{person}{Julian Alverio}, \bibinfo{person}{William Luo},
  \bibinfo{person}{Christopher Wang}, \bibinfo{person}{Dan Gutfreund},
  \bibinfo{person}{Josh Tenenbaum}, {and} \bibinfo{person}{Boris Katz}.}
  \bibinfo{year}{2019}\natexlab{}.
\newblock \showarticletitle{Objectnet: A large-scale bias-controlled dataset
  for pushing the limits of object recognition models}.
\newblock \bibinfo{journal}{\emph{Advances in neural information processing
  systems}}  \bibinfo{volume}{32} (\bibinfo{year}{2019}).
\newblock


\bibitem[Barron et~al\mbox{.}(2009)]%
        {barron2009parents}
\bibfield{author}{\bibinfo{person}{Brigid Barron},
  \bibinfo{person}{Caitlin~Kennedy Martin}, \bibinfo{person}{Lori Takeuchi},
  {and} \bibinfo{person}{Rachel Fithian}.} \bibinfo{year}{2009}\natexlab{}.
\newblock \showarticletitle{Parents as learning partners in the development of
  technological fluency}.
\newblock  (\bibinfo{year}{2009}).
\newblock


\bibitem[Bell(2015)]%
        {bell2015learning}
\bibfield{author}{\bibinfo{person}{Amanda~M Bell}.}
  \bibinfo{year}{2015}\natexlab{}.
\newblock \showarticletitle{Learning complex systems with story-building in
  scratch}. In \bibinfo{booktitle}{\emph{Proceedings of the 14th International
  Conference on Interaction Design and Children}}. \bibinfo{pages}{307--310}.
\newblock


\bibitem[Beneteau et~al\mbox{.}(2019)]%
        {beneteau2019communication}
\bibfield{author}{\bibinfo{person}{Erin Beneteau}, \bibinfo{person}{Olivia~K
  Richards}, \bibinfo{person}{Mingrui Zhang}, \bibinfo{person}{Julie~A Kientz},
  \bibinfo{person}{Jason Yip}, {and} \bibinfo{person}{Alexis Hiniker}.}
  \bibinfo{year}{2019}\natexlab{}.
\newblock \showarticletitle{Communication breakdowns between families and
  Alexa}. In \bibinfo{booktitle}{\emph{Proceedings of the 2019 CHI conference
  on human factors in computing systems}}. \bibinfo{pages}{1--13}.
\newblock


\bibitem[Berland and Reiser(2009)]%
        {berland2009making}
\bibfield{author}{\bibinfo{person}{Leema~Kuhn Berland} {and}
  \bibinfo{person}{Brian~J Reiser}.} \bibinfo{year}{2009}\natexlab{}.
\newblock \showarticletitle{Making sense of argumentation and explanation}.
\newblock \bibinfo{journal}{\emph{Science education}} \bibinfo{volume}{93},
  \bibinfo{number}{1} (\bibinfo{year}{2009}), \bibinfo{pages}{26--55}.
\newblock


\bibitem[Blackwell et~al\mbox{.}(2001)]%
        {blackwell2001cognitive}
\bibfield{author}{\bibinfo{person}{Alan~F Blackwell}, \bibinfo{person}{Carol
  Britton}, \bibinfo{person}{Anna Cox}, \bibinfo{person}{Thomas~RG Green},
  \bibinfo{person}{Corin Gurr}, \bibinfo{person}{Gada Kadoda},
  \bibinfo{person}{Maria~S Kutar}, \bibinfo{person}{Martin Loomes},
  \bibinfo{person}{Chrystopher~L Nehaniv}, \bibinfo{person}{Marian Petre},
  {et~al\mbox{.}}} \bibinfo{year}{2001}\natexlab{}.
\newblock \showarticletitle{Cognitive dimensions of notations: Design tools for
  cognitive technology}. In \bibinfo{booktitle}{\emph{International conference
  on cognitive technology}}. Springer, \bibinfo{pages}{325--341}.
\newblock


\bibitem[Blumenfeld et~al\mbox{.}(1996)]%
        {blumenfeld1996learning}
\bibfield{author}{\bibinfo{person}{Phyllis~C Blumenfeld},
  \bibinfo{person}{Ronald~W Marx}, \bibinfo{person}{Elliot Soloway}, {and}
  \bibinfo{person}{Joseph Krajcik}.} \bibinfo{year}{1996}\natexlab{}.
\newblock \showarticletitle{Learning with peers: From small group cooperation
  to collaborative communities}.
\newblock \bibinfo{journal}{\emph{Educational researcher}}
  \bibinfo{volume}{25}, \bibinfo{number}{8} (\bibinfo{year}{1996}),
  \bibinfo{pages}{37--39}.
\newblock


\bibitem[Buolamwini and Gebru(2018)]%
        {buolamwini2018gender}
\bibfield{author}{\bibinfo{person}{Joy Buolamwini} {and}
  \bibinfo{person}{Timnit Gebru}.} \bibinfo{year}{2018}\natexlab{}.
\newblock \showarticletitle{Gender shades: Intersectional accuracy disparities
  in commercial gender classification}. In \bibinfo{booktitle}{\emph{Conference
  on fairness, accountability and transparency}}. PMLR,
  \bibinfo{pages}{77--91}.
\newblock


\bibitem[Carney et~al\mbox{.}(2020)]%
        {carney2020teachable}
\bibfield{author}{\bibinfo{person}{Michelle Carney}, \bibinfo{person}{Barron
  Webster}, \bibinfo{person}{Irene Alvarado}, \bibinfo{person}{Kyle Phillips},
  \bibinfo{person}{Noura Howell}, \bibinfo{person}{Jordan Griffith},
  \bibinfo{person}{Jonas Jongejan}, \bibinfo{person}{Amit Pitaru}, {and}
  \bibinfo{person}{Alexander Chen}.} \bibinfo{year}{2020}\natexlab{}.
\newblock \showarticletitle{Teachable machine: Approachable Web-based tool for
  exploring machine learning classification}. In
  \bibinfo{booktitle}{\emph{Extended Abstracts of the 2020 CHI Conference on
  Human Factors in Computing Systems}}. \bibinfo{pages}{1--8}.
\newblock


\bibitem[Davis et~al\mbox{.}(2017)]%
        {davis2017haptic}
\bibfield{author}{\bibinfo{person}{Richard~Lee Davis}, \bibinfo{person}{Melisa
  Orta~Martinez}, \bibinfo{person}{Oliver Schneider}, \bibinfo{person}{Karon~E
  MacLean}, \bibinfo{person}{Allison~M Okamura}, {and} \bibinfo{person}{Paulo
  Blikstein}.} \bibinfo{year}{2017}\natexlab{}.
\newblock \showarticletitle{The haptic bridge: Towards a theory for
  haptic-supported learning}. In \bibinfo{booktitle}{\emph{Proceedings of the
  2017 conference on interaction design and children}}.
  \bibinfo{pages}{51--60}.
\newblock


\bibitem[Di~Blas and Boretti(2009)]%
        {di2009interactive}
\bibfield{author}{\bibinfo{person}{Nicoletta Di~Blas} {and}
  \bibinfo{person}{Bianca Boretti}.} \bibinfo{year}{2009}\natexlab{}.
\newblock \showarticletitle{Interactive storytelling in pre-school: a
  case-study}. In \bibinfo{booktitle}{\emph{Proceedings of the 8th
  International conference on interaction design and children}}.
  \bibinfo{pages}{44--51}.
\newblock


\bibitem[Druga(2018)]%
        {druga2018growing}
\bibfield{author}{\bibinfo{person}{Stefania Druga}.}
  \bibinfo{year}{2018}\natexlab{}.
\newblock \emph{\bibinfo{title}{Growing up with AI: Cognimates: from coding to
  teaching machines}}.
\newblock \bibinfo{thesistype}{Ph.\,D. Dissertation}.
  \bibinfo{school}{Massachusetts Institute of Technology}.
\newblock


\bibitem[Druga et~al\mbox{.}(2022)]%
        {druga2022family}
\bibfield{author}{\bibinfo{person}{Stefania Druga}, \bibinfo{person}{Fee~Lia
  Christoph}, {and} \bibinfo{person}{Amy~J Ko}.}
  \bibinfo{year}{2022}\natexlab{}.
\newblock \showarticletitle{Family as a Third Space for AI Literacies: How do
  children and parents learn about AI together?}. In
  \bibinfo{booktitle}{\emph{CHI Conference on Human Factors in Computing
  Systems}}. \bibinfo{pages}{1--17}.
\newblock


\bibitem[Dwivedi et~al\mbox{.}(2021)]%
        {dwivedi2021exploring}
\bibfield{author}{\bibinfo{person}{Utkarsh Dwivedi}, \bibinfo{person}{Jaina
  Gandhi}, \bibinfo{person}{Raj Parikh}, \bibinfo{person}{Merijke Coenraad},
  \bibinfo{person}{Elizabeth Bonsignore}, {and} \bibinfo{person}{Hernisa
  Kacorri}.} \bibinfo{year}{2021}\natexlab{}.
\newblock \showarticletitle{Exploring Machine Teaching with Children}. In
  \bibinfo{booktitle}{\emph{2021 IEEE Symposium on Visual Languages and
  Human-Centric Computing (VL/HCC)}}. IEEE, \bibinfo{pages}{1--11}.
\newblock


\bibitem[Eubanks(2018)]%
        {eubanks2018automating}
\bibfield{author}{\bibinfo{person}{Virginia Eubanks}.}
  \bibinfo{year}{2018}\natexlab{}.
\newblock \bibinfo{booktitle}{\emph{Automating inequality: How high-tech tools
  profile, police, and punish the poor}}.
\newblock \bibinfo{publisher}{St. Martin's Press}.
\newblock


\bibitem[Fiebrink and Cook(2010)]%
        {fiebrink2010wekinator}
\bibfield{author}{\bibinfo{person}{Rebecca Fiebrink} {and}
  \bibinfo{person}{Perry~R Cook}.} \bibinfo{year}{2010}\natexlab{}.
\newblock \showarticletitle{The Wekinator: a system for real-time, interactive
  machine learning in music}. In \bibinfo{booktitle}{\emph{Proceedings of The
  Eleventh International Society for Music Information Retrieval Conference
  (ISMIR 2010)(Utrecht)}}, Vol.~\bibinfo{volume}{3}.
\newblock


\bibitem[Gamelab(2003)]%
        {diner-dash}
\bibfield{author}{\bibinfo{person}{Gamelab}.} \bibinfo{year}{2003}\natexlab{}.
\newblock \bibinfo{title}{Diner Dash}.
\newblock
\newblock
\urldef\tempurl%
\url{https://www.crazygames.com/game/diner-dash}
\showURL{%
\tempurl}


\bibitem[Games(2016)]%
        {overcooked}
\bibfield{author}{\bibinfo{person}{Ghost~Town Games}.}
  \bibinfo{year}{2016}\natexlab{}.
\newblock \bibinfo{title}{Overcooked}.
\newblock
\newblock
\urldef\tempurl%
\url{https://www.team17.com/games/overcooked/}
\showURL{%
\tempurl}


\bibitem[Hohman et~al\mbox{.}(2020)]%
        {hohman2020understanding}
\bibfield{author}{\bibinfo{person}{Fred Hohman}, \bibinfo{person}{Kanit
  Wongsuphasawat}, \bibinfo{person}{Mary~Beth Kery}, {and}
  \bibinfo{person}{Kayur Patel}.} \bibinfo{year}{2020}\natexlab{}.
\newblock \showarticletitle{Understanding and Visualizing Data Iteration in
  Machine Learning}. In \bibinfo{booktitle}{\emph{Proceedings of the SIGCHI
  Conference on Human Factors in Computing Systems}}. ACM.
\newblock
\urldef\tempurl%
\url{https://doi.org/10.1145/3313831.3376177}
\showDOI{\tempurl}


\bibitem[Hong et~al\mbox{.}(2020)]%
        {hong2020crowdsourcing}
\bibfield{author}{\bibinfo{person}{Jonggi Hong}, \bibinfo{person}{Kyungjun
  Lee}, \bibinfo{person}{June Xu}, {and} \bibinfo{person}{Hernisa Kacorri}.}
  \bibinfo{year}{2020}\natexlab{}.
\newblock \showarticletitle{Crowdsourcing the perception of machine teaching}.
  In \bibinfo{booktitle}{\emph{Proceedings of the 2020 CHI Conference on Human
  Factors in Computing Systems}}. \bibinfo{pages}{1--14}.
\newblock


\bibitem[Kahn and Winters(2017)]%
        {kahn2017child}
\bibfield{author}{\bibinfo{person}{Ken Kahn} {and} \bibinfo{person}{Niall
  Winters}.} \bibinfo{year}{2017}\natexlab{}.
\newblock \showarticletitle{Child-friendly programming interfaces to AI cloud
  services}. In \bibinfo{booktitle}{\emph{European Conference on Technology
  Enhanced Learning}}. Springer, \bibinfo{pages}{566--570}.
\newblock


\bibitem[Lee et~al\mbox{.}(2021)]%
        {lee2021developing}
\bibfield{author}{\bibinfo{person}{Irene Lee}, \bibinfo{person}{Safinah Ali},
  \bibinfo{person}{Helen Zhang}, \bibinfo{person}{Daniella DiPaola}, {and}
  \bibinfo{person}{Cynthia Breazeal}.} \bibinfo{year}{2021}\natexlab{}.
\newblock \showarticletitle{Developing Middle School Students' AI Literacy}. In
  \bibinfo{booktitle}{\emph{Proceedings of the 52nd ACM Technical Symposium on
  Computer Science Education}}. \bibinfo{pages}{191--197}.
\newblock


\bibitem[Lin and Liu(2012)]%
        {lin2012investigation}
\bibfield{author}{\bibinfo{person}{Janet Mei-Chuen Lin} {and}
  \bibinfo{person}{Shu-Fen Liu}.} \bibinfo{year}{2012}\natexlab{}.
\newblock \showarticletitle{An investigation into parent-child collaboration in
  learning computer programming}.
\newblock \bibinfo{journal}{\emph{Journal of Educational Technology \&
  Society}} \bibinfo{volume}{15}, \bibinfo{number}{1} (\bibinfo{year}{2012}),
  \bibinfo{pages}{162--173}.
\newblock


\bibitem[Long and Magerko(2020)]%
        {long2020ai}
\bibfield{author}{\bibinfo{person}{Duri Long} {and} \bibinfo{person}{Brian
  Magerko}.} \bibinfo{year}{2020}\natexlab{}.
\newblock \showarticletitle{What is AI literacy? Competencies and design
  considerations}. In \bibinfo{booktitle}{\emph{Proceedings of the 2020 CHI
  Conference on Human Factors in Computing Systems}}. \bibinfo{pages}{1--16}.
\newblock


\bibitem[Long et~al\mbox{.}(2021)]%
        {long2021role}
\bibfield{author}{\bibinfo{person}{Duri Long}, \bibinfo{person}{Aadarsh
  Padiyath}, \bibinfo{person}{Anthony Teachey}, {and} \bibinfo{person}{Brian
  Magerko}.} \bibinfo{year}{2021}\natexlab{}.
\newblock \showarticletitle{The Role of Collaboration, Creativity, and
  Embodiment in AI Learning Experiences}. In
  \bibinfo{booktitle}{\emph{Creativity and Cognition}}. \bibinfo{pages}{1--10}.
\newblock


\bibitem[Long et~al\mbox{.}(2022)]%
        {long2022family}
\bibfield{author}{\bibinfo{person}{Duri Long}, \bibinfo{person}{Anthony
  Teachey}, {and} \bibinfo{person}{Brian Magerko}.}
  \bibinfo{year}{2022}\natexlab{}.
\newblock \showarticletitle{Family Learning Talk in AI Literacy Learning
  Activities}. In \bibinfo{booktitle}{\emph{CHI Conference on Human Factors in
  Computing Systems}}. \bibinfo{pages}{1--20}.
\newblock


\bibitem[Lyons et~al\mbox{.}(2015)]%
        {lyons2015designing}
\bibfield{author}{\bibinfo{person}{Leilah Lyons}, \bibinfo{person}{Michael
  Tissenbaum}, \bibinfo{person}{Matthew Berland}, \bibinfo{person}{Rebecca
  Eydt}, \bibinfo{person}{Lauren Wielgus}, {and} \bibinfo{person}{Adam
  Mechtley}.} \bibinfo{year}{2015}\natexlab{}.
\newblock \showarticletitle{Designing visible engineering: supporting tinkering
  performances in museums}. In \bibinfo{booktitle}{\emph{Proceedings of the
  14th International Conference on Interaction Design and Children}}.
  \bibinfo{pages}{49--58}.
\newblock


\bibitem[Noble(2018)]%
        {noble2018algorithms}
\bibfield{author}{\bibinfo{person}{Safiya~Umoja Noble}.}
  \bibinfo{year}{2018}\natexlab{}.
\newblock \bibinfo{booktitle}{\emph{Algorithms of oppression}}.
\newblock \bibinfo{publisher}{New York University Press}.
\newblock


\bibitem[Payne(2019)]%
        {payne2019ethics}
\bibfield{author}{\bibinfo{person}{Blakeley~H Payne}.}
  \bibinfo{year}{2019}\natexlab{}.
\newblock \showarticletitle{An ethics of artificial intelligence curriculum for
  middle school students}.
\newblock  (\bibinfo{year}{2019}).
\newblock


\bibitem[Roque et~al\mbox{.}(2016)]%
        {roque2016m}
\bibfield{author}{\bibinfo{person}{Ricarose Roque}, \bibinfo{person}{Karina
  Lin}, {and} \bibinfo{person}{Richard Liuzzi}.}
  \bibinfo{year}{2016}\natexlab{}.
\newblock \showarticletitle{“I’m Not Just a Mom”: Parents Developing
  Multiple Roles in Creative Computing}.
\newblock \bibinfo{publisher}{Singapore: International Society of the Learning
  Sciences}.
\newblock


\bibitem[Roschelle(1992)]%
        {roschelle1992learning}
\bibfield{author}{\bibinfo{person}{Jeremy Roschelle}.}
  \bibinfo{year}{1992}\natexlab{}.
\newblock \showarticletitle{Learning by collaborating: Convergent conceptual
  change}.
\newblock \bibinfo{journal}{\emph{The journal of the learning sciences}}
  \bibinfo{volume}{2}, \bibinfo{number}{3} (\bibinfo{year}{1992}),
  \bibinfo{pages}{235--276}.
\newblock


\bibitem[Sculley et~al\mbox{.}(2015)]%
        {sculley2015hidden}
\bibfield{author}{\bibinfo{person}{David Sculley}, \bibinfo{person}{Gary Holt},
  \bibinfo{person}{Daniel Golovin}, \bibinfo{person}{Eugene Davydov},
  \bibinfo{person}{Todd Phillips}, \bibinfo{person}{Dietmar Ebner},
  \bibinfo{person}{Vinay Chaudhary}, \bibinfo{person}{Michael Young},
  \bibinfo{person}{Jean-Francois Crespo}, {and} \bibinfo{person}{Dan
  Dennison}.} \bibinfo{year}{2015}\natexlab{}.
\newblock \showarticletitle{Hidden technical debt in machine learning systems}.
\newblock \bibinfo{journal}{\emph{Advances in neural information processing
  systems}}  \bibinfo{volume}{28} (\bibinfo{year}{2015}).
\newblock


\bibitem[Touretzky et~al\mbox{.}(2019)]%
        {ai4k12}
\bibfield{author}{\bibinfo{person}{David~S. Touretzky},
  \bibinfo{person}{Christina Gardner-McCune}, \bibinfo{person}{Fred Martin},
  {and} \bibinfo{person}{Deborah Seehorn}.} \bibinfo{year}{2019}\natexlab{}.
\newblock \bibinfo{title}{K-12 Guidelines for Artificial Intelligence: What
  Students Should Know}.
\newblock
\newblock
\urldef\tempurl%
\url{https://github.com/touretzkyds/ai4k12/raw/master/documents/ISTE_2019_Presentation_website_final.pdf/}
\showURL{%
\tempurl}


\bibitem[Tseng et~al\mbox{.}(2021)]%
        {tseng2021plushpal}
\bibfield{author}{\bibinfo{person}{Tiffany Tseng}, \bibinfo{person}{Yumiko
  Murai}, \bibinfo{person}{Natalie Freed}, \bibinfo{person}{Deanna Gelosi},
  \bibinfo{person}{Tung~D Ta}, {and} \bibinfo{person}{Yoshihiro Kawahara}.}
  \bibinfo{year}{2021}\natexlab{}.
\newblock \showarticletitle{PlushPal: Storytelling with Interactive Plush Toys
  and Machine Learning}. In \bibinfo{booktitle}{\emph{Interaction Design and
  Children}}. \bibinfo{pages}{236--245}.
\newblock


\bibitem[von Wangenheim et~al\mbox{.}(2021)]%
        {von2021visual}
\bibfield{author}{\bibinfo{person}{Christiane~Gresse von Wangenheim},
  \bibinfo{person}{Jean~CR Hauck}, \bibinfo{person}{Fernando~S Pacheco}, {and}
  \bibinfo{person}{Matheus F~Bertonceli Bueno}.}
  \bibinfo{year}{2021}\natexlab{}.
\newblock \showarticletitle{Visual tools for teaching machine learning in K-12:
  A ten-year systematic mapping}.
\newblock \bibinfo{journal}{\emph{Education and Information Technologies}}
  (\bibinfo{year}{2021}), \bibinfo{pages}{1--46}.
\newblock


\bibitem[Vygotsky and Cole(1978)]%
        {vygotsky1978mind}
\bibfield{author}{\bibinfo{person}{Lev~Semenovich Vygotsky} {and}
  \bibinfo{person}{Michael Cole}.} \bibinfo{year}{1978}\natexlab{}.
\newblock \bibinfo{booktitle}{\emph{Mind in society: Development of higher
  psychological processes}}.
\newblock \bibinfo{publisher}{Harvard university press}.
\newblock


\bibitem[Williams and Breazeal(2020)]%
        {williams2020train}
\bibfield{author}{\bibinfo{person}{Randi Williams} {and}
  \bibinfo{person}{Cynthia Breazeal}.} \bibinfo{year}{2020}\natexlab{}.
\newblock \showarticletitle{How to Train Your Robot: A Middle School AI and
  Ethics Curriculum}. IJCAI.
\newblock


\bibitem[Yang et~al\mbox{.}(2018)]%
        {yang2018grounding}
\bibfield{author}{\bibinfo{person}{Qian Yang}, \bibinfo{person}{Jina Suh},
  \bibinfo{person}{Nan-Chen Chen}, {and} \bibinfo{person}{Gonzalo Ramos}.}
  \bibinfo{year}{2018}\natexlab{}.
\newblock \showarticletitle{Grounding interactive machine learning tool design
  in how non-experts actually build models}. In
  \bibinfo{booktitle}{\emph{Proceedings of the 2018 designing interactive
  systems conference}}. \bibinfo{pages}{573--584}.
\newblock


\bibitem[Yin(2009)]%
        {yin2009case}
\bibfield{author}{\bibinfo{person}{Robert~K Yin}.}
  \bibinfo{year}{2009}\natexlab{}.
\newblock \bibinfo{booktitle}{\emph{Case study research: Design and methods}}.
  Vol.~\bibinfo{volume}{5}.
\newblock \bibinfo{publisher}{sage}.
\newblock


\bibitem[Yip et~al\mbox{.}(2014)]%
        {yip2014helped}
\bibfield{author}{\bibinfo{person}{Jason Yip}, \bibinfo{person}{June Ahn},
  \bibinfo{person}{Tamara Clegg}, \bibinfo{person}{Elizabeth Bonsignore},
  \bibinfo{person}{Daniel Pauw}, {and} \bibinfo{person}{Michael Gubbels}.}
  \bibinfo{year}{2014}\natexlab{}.
\newblock \showarticletitle{" It helped me do my science." a case of designing
  social media technologies for children in science learning}. In
  \bibinfo{booktitle}{\emph{Proceedings of the 2014 conference on Interaction
  design and children}}. \bibinfo{pages}{155--164}.
\newblock


\bibitem[Zhou et~al\mbox{.}(2020)]%
        {zhou2020designing}
\bibfield{author}{\bibinfo{person}{Xiaofei Zhou}, \bibinfo{person}{Jessica
  Van~Brummelen}, {and} \bibinfo{person}{Phoebe Lin}.}
  \bibinfo{year}{2020}\natexlab{}.
\newblock \showarticletitle{Designing AI Learning Experiences for K-12:
  Emerging Works, Future Opportunities and a Design Framework}.
\newblock \bibinfo{journal}{\emph{arXiv preprint arXiv:2009.10228}}
  (\bibinfo{year}{2020}).
\newblock


\bibitem[Zimmerman et~al\mbox{.}(2010)]%
        {zimmerman2010parents}
\bibfield{author}{\bibinfo{person}{Heather Zimmerman}, \bibinfo{person}{Suzanne
  Perin}, {and} \bibinfo{person}{Philip Bell}.}
  \bibinfo{year}{2010}\natexlab{}.
\newblock \showarticletitle{Parents, science, and interest}.
\newblock \bibinfo{journal}{\emph{Museums \& Social Issues}}
  \bibinfo{volume}{5}, \bibinfo{number}{1} (\bibinfo{year}{2010}),
  \bibinfo{pages}{67--86}.
\newblock


\bibitem[Zimmermann-Niefield et~al\mbox{.}(2020)]%
        {zimmermann2020youth}
\bibfield{author}{\bibinfo{person}{Abigail Zimmermann-Niefield},
  \bibinfo{person}{Shawn Polson}, \bibinfo{person}{Celeste Moreno}, {and}
  \bibinfo{person}{R~Benjamin Shapiro}.} \bibinfo{year}{2020}\natexlab{}.
\newblock \showarticletitle{Youth making machine learning models for
  gesture-controlled interactive media}. In
  \bibinfo{booktitle}{\emph{Proceedings of the Interaction Design and Children
  Conference}}. \bibinfo{pages}{63--74}.
\newblock


\bibitem[Zimmermann-Niefield et~al\mbox{.}(2019)]%
        {zimmermann2019youth}
\bibfield{author}{\bibinfo{person}{Abigail Zimmermann-Niefield},
  \bibinfo{person}{Makenna Turner}, \bibinfo{person}{Bridget Murphy},
  \bibinfo{person}{Shaun~K Kane}, {and} \bibinfo{person}{R~Benjamin Shapiro}.}
  \bibinfo{year}{2019}\natexlab{}.
\newblock \showarticletitle{Youth learning machine learning through building
  models of athletic moves}. In \bibinfo{booktitle}{\emph{Proceedings of the
  18th ACM International Conference on Interaction Design and Children}}.
  \bibinfo{pages}{121--132}.
\newblock


\end{thebibliography}

% \appendix

\end{document}